# NONLINEAR EVOLUTION OF GENUS
# IN PRIMORDIAL RANDOM – GAUSSIAN DENSITY FIELD

Takahiko Matsubara and Yasushi Suto[1]

*Department of Physics, The University of Tokyo, Tokyo 113, Japan.*

e-mail: matsu@utaphp1.phys.s.u-tokyo.ac.jp,   suto@phys.s.u-tokyo.ac.jp

## ABSTRACT

The genus statistics is studied using large $N$-body simulations for several cosmological models. We consider the effects of nonlinear gravitational evolution, smoothing the particle data in fully nonlinear regime, and the redshift-space distortion on the genus curve. Detailed comparison between the theoretical prediction in weakly nonlinear theory and the appropriate simulation results shows that the analytic formula describes the behavior of genus in weakly nonlinear regime fairly accurately. We also find that the redshift-space distortion on genus statistics is small in linear and weakly nonlinear regimes. We conclude that if weakly nonlinear theory and direct numerical simulations are combined, the normalized genus curve $G(\nu)/G(0)$ is a powerful tool to directly check the random-Gaussian paradigm of the origin of the large-scale structure in the universe.

*Subject headings*: cosmology: theory — galaxies: formation — gravitation — methods: numerical

## 1.  INTRODUCTION

Since the pioneering work of Gott, Melott, & Dickinson (1986, GMD), topological analysis on the basis of the genus has been applied in statistical description of the galaxy clustering by various authors (Gott, Weinberg & Melott 1987; Weinberg, Gott & Melott 1987; Melott, Weinberg & Gott 1988; Gott et al. 1989; Park & Gott 1991; Park, Gott & da Costa 1992; Weinberg & Cole 1992; Moore et al. 1992; Vogeley et al. 1994; Rhoads, Gott & Postman 1994). When one denotes the Fourier transform of the density field (obtained from the spatial distribution of galaxies, assuming that galaxies trace mass) by $\delta(k) = |\delta(k)|\exp(i\phi(k))$, the two-point correlation $\xi(r)$, a more conventional measure of the galaxy clustering, is defined only through the amplitude $|\delta(k)|$, or its ensemble average $P(k) \equiv \langle|\delta(k)|^2\rangle$. In contrast, the genus statistics $G(\nu)$ depends on the phase $\phi(k)$ as well as $|\delta(k)|$, and thus quantifies information of the galaxy clustering which cannot be described in terms of $\xi(r)$. In particular, genus statistics is suited to test the random-Gaussianity of the primordial density field (Hamilton, Gott, & Weinberg 1986).

A practical limitation in performing the test with the observed data comes from the size of the sample itself. The random-Gaussian nature of the primordial density field is lost due to the nonlinear gravitational evolution, and can be recovered only by smoothing the *present* galaxy distribution over a scale much larger than $10h^{-1}\mathrm{Mpc}$. Since the effective volume of the currently available surveys in three dimension is $\sim (100h^{-1}\mathrm{Mpc})^3$ at most, such a large smoothing significantly reduces the number of statistically independent samples. This difficulty is removed if one can correct for the nonlinear effect with theory; then one may apply a smaller smoothing length to the limited observed samples, and thus improve the statistics. In this respect, the analytic expression for genus in a weakly nonlinear regime from primordial random-Gaussian density field (Matsubara 1994b) is of great value in quantitative comparison between cosmological models and the observations. In particular this formula significantly increases the possibility of testing the primordial random-Gaussianity with the current and future galaxy redshift surveys.

Another possible limitation of the galaxy redshift surveys may result from the redshift-space contamination. The proper mapping of the observed data in redshift space to those in real space requires the information of the peculiar velocity field, which is usually complicated and difficult to obtain in practice. In fact, it is shown that the two-point correlation functions suffer from the significant redshift-space contamination (Davis & Peebles 1983; Lilje & Efstathiou 1989; Suto & Suginohara 1991; Matsubara 1994a). Similar effect is predicted also on the higher-order statistics (Lahav et al. 1993; Matsubara & Suto 1994; Suto & Matsubara 1994; Ghigna et al. 1994; Fry & Gaztañaga 1994). Therefore it is important to examine the extent to which the genus statistics is affected by the peculiar velocity field. Matsubara (1996) showed that the redshift-space contamination in linear theory does not change the shape of the genus itself, but suppresses the overall amplitude. Thus it is interesting to see how the tendency changes in a strongly nonlinear regime.

This paper examines the nonlinear behavior and redshift-space contamination of the the genus statistics for several models computed with cosmological $N$-body simulations (Suginohara et al. 1991; Suto 1993). The simulation results are compared with with the theoretical predictions in linear and weakly nonlinear regimes mentioned-above (Matsubara 1994b, 1996). The plan of the paper is as follows; in §2 we briefly summarize several analytical results related to the genus statistics. The simulation models and the analysis method are described in §3. Then we present results of the nonlinear effect, redshift-space contamination, and the smoothing effect in order. Finally §4 is devoted to pur conclusions and further discussion.

## 2. ANALYTIC PREDICTIONS

### 2.1. Random-Gaussian model

The Gauss-Bonnet theorem states that the genus $g$ defined in terms of the Gaussian curvature $\kappa$ of a compact two-dimensional surface $A$:

$$g \equiv -\frac{1}{4\pi} \int \kappa \, dA \tag{2.1}$$

is simply (*the number of the holes of the surface*) $-1$. If there are more than one two-dimensional surface $A_i$ ($i = 1 \sim I$) in the volume $V$, one may define the *genus density $G$* as

$$G \equiv \frac{1}{V} \sum_{i=1}^{I} g_i = -\frac{1}{4\pi V} \sum_{i=1}^{I} \int_{A_i} \kappa \, dA. \tag{2.2}$$

GMD showed that the genus density of the isodensity surface of the cosmological density field can be a useful statistics to characterize the topology of the large-scale structure in the universe. In this context it is conventional (Hamilton, Gott & Weinberg 1986) to define the genus density with the isodensity threshold $\nu$ as the *genus curve $G(\nu)$*. The corresponding isodensity surfaces are where the density fluctuations $\delta$ has the value $\nu\sigma$ with $\sigma$ being the rms value of the fluctuations.

Fluctuations on large scales, which are in linear regime, are supposed to retain the Gaussianity if primordial fluctuations are random-Gaussian field (but see §3.4 below). In this case the genus curve can be computed analytically (Adler 1981; Doroshkevich 1970; Bardeen et al. 1986; and Hamilton et al. 1986), and is given by

$$G(\nu) = \frac{1}{4\pi^2} \left( \frac{\langle k^2 \rangle}{3} \right)^{3/2} e^{-\nu^2/2}(1 - \nu^2). \tag{2.3}$$

Thus the amplitude of the genus curve for random Gaussian field is characterized by

$$\langle k^2 \rangle \equiv \frac{\int k^2 P(k) W^2(kR) d^3k}{\int P(k) W^2(kR) d^3k}, \tag{2.4}$$

where $P(k)$ is the power spectrum of the density fluctuation, and $W(kR)$ is a window function to smooth the underlying density field. Throughout the present analysis we use a Gaussian window $W(kR) = \exp(-k^2 R^2/2)$ with $R$ being the corresponding smoothing (or filtering) length. The genus curve for random-Gaussian field (2.3) is positive for $|\nu| \leq 1$, negative for $|\nu| \geq 1$, and approaches zero from negative value as $|\nu|$ increases. These features can be easily understood; the number of holes exceeds that of isolated regions for smaller $|\nu|$ where density threshold is around the mean density, and therefore the topology of the surfaces is sponge-like. On the contrary, the number of isolated regions dominates that of holes for larger $|\nu|$, and the topology becomes meatball-like (GMD). As $|\nu|$ increases further, the number of such rare objects rapidly decreases. These qualitative features are quite generic and, as we will show below, are not significantly affected by the gravitational nonlinear evolution.

## 2.2.  Effect of weakly nonlinear gravitational evolution

Since the observed density field probed by the galaxy distribution has inevitably experienced the nonlinear gravitational evolution, one cannot compare the observational data directly with the random $-$ Gaussian prediction (2.3). Matsubara (1994b) takes into account the effect of the nonlinearity on the genus curve using perturbation theory. We first summarize his analytic formula for $G(\nu)$ in a weakly nonlinear regime. Then we present specific expressions in models with arbitrary density parameter $\Omega$ and the dimensionless cosmological constant $\lambda$ by combining with the second-order perturbation theory (Bouchet et al. 1992; Bernardeau 1994; Matsubara 1995).

The irreducible or connected parts of higher order correlations $\langle \delta^n \rangle_c$ ($n \geq 3$) (e.g., Bertchinger 1992), which vanish in random $-$ Gaussian fields, do not vanish for general non-Gaussian fields. However if they satisfy the following:

$$\langle \delta^3 \rangle \sim \mathcal{O}(\sigma^4),  \tag{2.5}$$

$$\langle \delta^n \rangle_c \sim \mathcal{O}(\sigma^{n+2}) \text{ or higher } (n \geq 4),  \tag{2.6}$$

where $\sigma$ is the rms amplitude of the density fluctuations, then the genus curve in such a field is generally given by (Matsubara 1994b)

$$G(\nu) = -\frac{1}{4\pi^2} \left( \frac{\langle k^2 \rangle}{3} \right)^{3/2} e^{-\nu^2/2} \left[ H_2(\nu) + \sigma \left( \frac{S}{6} H_5(\nu) + \frac{3T}{2} H_3(\nu) + 3U H_1(\nu) \right) + \mathcal{O}(\sigma^2) \right].  \tag{2.7}$$

In the above expression, $H_n(\nu) \equiv (-)^n e^{\nu^2/2} (d/d\nu)^n e^{-\nu^2/2}$ are the Hermite polynomials, and $S$, $T$, and $U$ are defined as

$$
\begin{aligned}
S &= \frac{1}{\sigma^4} \langle \delta^3 \rangle, \\
T &= -\frac{1}{2\langle k^2 \rangle \sigma^4} \langle \delta^2 \nabla^2 \delta \rangle, \\
U &= -\frac{3}{4\langle k^2 \rangle^2 \sigma^4} \langle \nabla \delta \cdot \nabla \delta \nabla^2 \delta \rangle.
\end{aligned}
\tag{2.8}
$$

In fact the conditions (2.5) and (2.6) are satisfied in a weakly nonlinear cosmological density field from primordial random-Gaussian field; higher-order perturbation analysis (Fry 1984; Goroff et al. 1986; Bernardeau 1992) predicts the following hierarchical relation among correlations:

$$\langle \delta^n \rangle_c \sim \mathcal{O}(\sigma^{2n-2}) \ (n \geq 2).  \tag{2.9}$$

Therefore we can use the general formula (2.7) to compute the genus curve in such weakly nonlinear regimes. The generalized skewness $S$, $T$ and $U$ can be evaluated also by perturbation theory (Matsubara 1994b). If one uses the Gaussian window with the smoothing length $R$, they are explicitly computed as

$$S = \frac{1}{4\pi^4}\left[(2+K)L_{220} + 3L_{131} + (1-K)L_{222}\right],$$

$$T = \frac{1}{60\pi^4}[5(5+2K)L_{240} + 3(9+K)L_{331} + 15L_{151}$$
$$+ 10(2-K)L_{242} + 3(1-K)L_{333}], \tag{2.10}$$

$$U = \frac{1}{140\pi^4}\left[7(3+2K)L_{440} + 21L_{351} - 5(3+4K)L_{442} - 21L_{353} - 6(1-K)L_{444}\right].$$

Here we introduce the following integrals:

$$L_{\alpha\beta n}(R) \equiv \frac{\langle k^2\rangle^{2-(\alpha+\beta)/2}}{\sigma_R^4}\int_0^\infty dx \int_0^\infty dy \int_{-1}^1 d\mu\, e^{-R^2(x^2+y^2+\mu xy)} x^\alpha y^\beta P_n(\mu) P(x) P(y) \tag{2.11}$$

$$= (-)^n \sqrt{2\pi}\frac{\langle k^2\rangle^{2-(\alpha+\beta)/2}}{\sigma_R^4 R}\int_0^\infty dx \int_0^\infty dy\, e^{-R^2(x^2+y^2)} x^{\alpha-1/2} y^{\beta-1/2} I_{n+1/2}(xyR^2) P(x) P(y), \tag{2.12}$$

where $\sigma_R$ is the rms amplitude of the Gaussian smoothed density fluctuation with $R$, $P_n$ is the Legendre polynomial, and $I_\nu$ is a modified Bessel function.

The above results (2.10) to (2.12) hold for arbitrary cosmological models with $\Omega$ and $\lambda$. The latter effect manifests only through $K = K(\Omega, \lambda)$ which very weakly depends on $\Omega$ and $\lambda$ (Bouchet et al. 1992; Bernardeau 1994). The explicit form for $K$ is derived by Matsubara (1995) as

$$K(\Omega, \lambda) = \frac{\Omega}{4} - \frac{\lambda}{2} - \left(\int_0^1 dx\, X^{-3/2}\right)^{-1} + \frac{3}{2}\left(\int_0^1 dx\, X^{-3/2}\right)^{-2}\int_0^1 dx\, X^{-5/2}, \tag{2.13}$$

where

$$X(x) \equiv \Omega/x + \lambda x^2 + 1 - \Omega - \lambda. \tag{2.14}$$

In two special models of our interest, $K(1,0) = 3/7 = 0.4286$ and $K(0.2, 0.8) = 0.4335$.

For the power-law fluctuation spectra $P(k) \propto k^n$, $S$, $T$ and $U$ can be written down explicitly in terms of the hypergeometric function as

$$S = 3F\left(\frac{n+3}{2}, \frac{n+3}{2}, \frac{3}{2}; \frac{1}{4}\right) - (n+2-2K)F\left(\frac{n+3}{2}, \frac{n+3}{2}, \frac{5}{2}; \frac{1}{4}\right),$$

$$T = 3F\left(\frac{n+3}{2}, \frac{n+5}{2}, \frac{3}{2}; \frac{1}{4}\right) - (n+3-K)F\left(\frac{n+3}{2}, \frac{n+5}{2}, \frac{5}{2}; \frac{1}{4}\right)$$
$$+ \frac{(n-2)(1-K)}{15}F\left(\frac{n+3}{2}, \frac{n+5}{2}, \frac{7}{2}; \frac{1}{4}\right), \tag{2.15}$$

$$U = F\left(\frac{n+5}{2}, \frac{n+5}{2}, \frac{5}{2}; \frac{1}{4}\right) - \frac{n+4-4K}{5}F\left(\frac{n+5}{2}, \frac{n+5}{2}, \frac{7}{2}; \frac{1}{4}\right).$$

The expressions for $S$ in equations (2.10) and (2.15) are derived by Łokas et al. (1994) which are equivalent to the other form independently derived by Matsubara (1994b). Similarly we transform the expressions for $T$ and $U$ presented in Matsubara (1994b; eqs. [16] and [18]) using the function $L_{\alpha\beta n}(R)$, which are given in equations (2.10) and (2.15).

### 2.3.   Redshift-space distortion in linear theory

As described in Introduction, peculiar velocity field may significantly distort the shape and amplitude of the genus curve computed in redshift space. Matsubara (1996) proved that in a linear regime the genus curves for primordial random-Gaussian fluctuations have the same shape as functions of $\nu$ while their overall amplitudes are different. More specifically, the genus curve in redshift space, $G^{(s)}(\nu)$, is related to its real space counterpart $G^{(r)}(\nu)$ as

$$\frac{G^{(s)}(\nu)}{G^{(r)}(\nu)} = \frac{3\sqrt{3}}{2}\sqrt{u}(1-u), \tag{2.16}$$

where

$$u = \frac{1}{3}\frac{1 + \dfrac{6}{5}f + \dfrac{3}{7}f^2}{1 + \dfrac{2}{3}f + \dfrac{1}{5}f^2}. \tag{2.17}$$

The function $f$ is the logarithmic derivative of the linear growth rate $D(t)$ with respect to the scale factor $a$ and given as follows :

$$f(\Omega, \lambda) \equiv \frac{d\ln D}{d\ln a} = -1 - \Omega/2 + \lambda + \left(\int_0^1 dx\, X^{-3/2}\right)^{-1} \tag{2.18}$$

$$\sim \Omega^{0.6} + \frac{\lambda}{70}(1 + \frac{\Omega}{2}). \tag{2.19}$$

The empirical fitting formula in the second line is derived by Lahav et al. (1991), which implies that $f$ is approximately given by $\Omega^{0.6}$ (Peebles 1980) and $\lambda$-dependence is quite weak for parameters of our interest. In this case equation (2.16) is close to unity and redshift space contamination is negligible; for $(\Omega, \lambda) = (1,0)$ and $(0.2, 0.8)$, values of equation (2.16) are 0.944 and 0.987, respectively. Therefore it is interesting to see to what extent this insensitivity to the redshift-space contamination in linear theory changes in weakly nonlinear and fully nonlinear regimes.

## 3.   GENUS ANALYSIS OF NUMERICAL SIMULATIONS

### 3.1.   Simulation models and method of analysis

The analysis below is based on the four data sets from cosmological $N$-body simulations with random-Gaussian initial conditions. Three models are evolved in the Einstein-de Sitter universe with the scale-free initial fluctuation spectra (at expansion factor $a = 1.0$):

$$P(k) \propto k^n \qquad (n = -1, 0, \text{ and } 1). \tag{3.1}$$

The last model corresponds to a spatially-flat low-density cold dark matter (LCDM) model. In this specific example, we assume $\Omega_0 = 0.2$, $\lambda_0 = 0.8$, and $h = 1.0$ (Suginohara & Suto 1991). The amplitude of the power spectrum in the LCDM model at $a = 6$ is normalized so that the top-hat smoothed rms mass fluctuation is unity at $8h^{-1}$Mpc. In fact this LCDM model can be regarded to represent a specific example of the most successful cosmological scenarios so far (e.g., Suto 1993). All models are evolved with a hierarchical tree code implementing the fully periodic boundary condition in a cubic volume of $L^3$. The physical comoving size of the computational box in the LCDM model is $L = 100h^{-1}$Mpc. The number of particles employed in the simulations is $N = 64^3$, and the gravitational softening length is $\epsilon_g = L/1280$ in comoving. Further details of the simulation models and other extensive analyses are described in Hernquist, Bouchet & Suto (1991), Suginohara et al. (1991), Suginohara & Suto (1991), Suto (1993), Matsubara & Suto (1994), and, Suto & Matsubara (1994).

The computation of the genus from the particle data is performed using the code kindly provided by David Weinberg (Weinberg 1988; Gott et al. 1989). In short the procedure goes as follows; (i) the computational box is divided into $N_c^3 (= 128^3)$ cubes, and the density $\rho_g(\mathbf{r})$ at the center of each cell is computed using Cloud-In-Cell density assignment. (ii) the Fourier-transform:

$$\tilde{\rho}_g(\mathbf{k}) \equiv \frac{1}{L^3} \int \rho_g(\mathbf{r})\exp(i\mathbf{k}\cdot\mathbf{r})d^3r, \tag{3.2}$$

is convolved with the Gaussian filter, and transformed back to define a *smoothed* density of each cell (with the filtering length $R_f$):

$$\rho_s(\mathbf{r}; R_f) = \frac{L^3}{8\pi^3} \int \tilde{\rho}_g(\mathbf{k})\exp(-k^2 R_f^2/2 - i\mathbf{k}\cdot\mathbf{r})d^3k. \tag{3.3}$$

(iii) the rms amplitude of the density fluctuations is computed directly from the smoothed density:

$$\sigma(R_f) \equiv \sqrt{\langle(\rho_s/\bar{\rho}-1)^2\rangle}, \tag{3.4}$$

where $\bar{\rho}$ is the mean density of the particles. (iv) The isodensity surface of the critical density:

$$\rho_c \equiv [1 + \nu\sigma(R_f)]\,\bar{\rho} \tag{3.5}$$

is approximated by the boundary surface of the high-density ($\rho_s > \rho_c$) and low-density ($\rho_s < \rho_c$) cells. (v) Then the genus of the surface is computed by summing up the angle deficit $D(i,j,k)$ at the vertex of cell $(i,j,k)$ (cf., eq. [2.1]):

$$g_s(\nu) = -\frac{1}{4\pi} \sum_{i,j,k=1}^{N_c} D(i,j,k). \tag{3.6}$$

The way to compute $D(i,j,k)$ is detailed in Gott et al. (1986). The *genus curve* $G(\nu)$ is defined to be the number of genus per unit volume as a function of the threshold $\nu$. (vi) We repeated the above procedure 50 times using the bootstrap resampling method in order to estimate the statistical errors of $G(\nu)$.

It should be noted that earlier papers (e.g., Gott et al. 1989; Rhoads, Gott, & Postman 1994; Vogeley et al. 1994) *defined* $\rho_c(\nu)$ so that the volume fraction on the high-density side of the isodensity surface is equal to

$$f = \frac{1}{\sqrt{2\pi}} \int_\nu^\infty e^{-t^2/2} dt. \tag{3.7}$$

Both methods yield equivalent value for $\nu$ only if the density field is strictly random-Gaussian. Since we are interested in the departure from the primordial Gaussianity, we decided to adopt the straightforward definition (3.5) throughout the present analysis. As a matter of fact, the previous authors *intentionally* avoided to use equation (3.5); the contours of fixed volume fraction partly compensate the nonlinear gravitational evolution, and may be suitable to examine the topology without explicit knowledge of evolution of the one-point density probability (Vogeley et al. 1994). On the other hand, now we know the effect of gravitational evolution explicitly from the formula (2.7) at least in a weakly nonlinear regime. That is why our present analysis is based on the straightforward definition (3.5) unlike the previous work.

### 3.2.  *Non-Gaussian signature due to nonlinear gravitational evolution*

Since we are interested in the *shape* of the genus curve, we factor out the overall amplitude of $G(\nu)$ which is proportional to the second moment of the power spectrum (eq.[2.4]). To be more specific, we define the *normalized genus curve* by $G^{(r)}(\nu)/G^{(r)}(0)$ in real space and $G^{(s)}(\nu)/G^{(r)}(0)$ in redshift space. This procedure is suitable for our purpose; while each realization in simulation models would inevitably differ from the theoretical fluctuation power spectrum to some extent, we know that this effect can be completely separated out by the above normalization at least in a weakly nonlinear regime (see, eq.[2.7]). In practice, we first compute $G(\nu)$ at 51 bins (in equal interval) for $-3 \leq \nu \leq 3$. Then we estimate the amplitude of $G^{(r)}(0)$ by $\chi^2$-fitting the 7 data points around $\nu = 0$ to the weakly nonlinear formula (2.7) so that thus computed value of $G^{(r)}(0)$ is less affected by the statistical fluctuation at one data point. This procedure works in a weakly nonlinear regime, but the overall normalization of the resulting curves may be somewhat arbitrary as the nonlinearity increases where the formula (2.7) is no longer valid. This should be remembered in the comparison below.

The normalized genus curves $G(\nu)/G(0)$ are plotted in Figures 1 to 3 for power-law models with $n = 1$, 0, and $-1$, respectively. We select three different sets of the expansion factor $a$ (= 1 at the initial epoch) and the filtering length $R_f$ for each model so that the resulting $\sigma(R_f)$ covers weakly, fairly and fully nonlinear regimes (in upper, middle, and lower panels, respectively). The mean values of $G(\nu)/G(0)$ from the 50 bootstrap resampling analyses in real space are plotted in filled circles with the corresponding $1\sigma$ statistical error. Open triangles indicate the results in redshift space which will be discussed in the next subsection. For comparison we plot the weakly nonlinear formula (eqs.[2.7] to [2.15]) in solid curves, together with the random-Gaussian prediction (2.3) in dotted curves. As expected, our simulation results and the weakly nonlinear formula agree quite well (middle panels in Figs.1 to 3). In fact, the simulation results are in reasonable agreement when $-0.2 \lesssim \nu\sigma \lesssim 0.4$ for all models, even though the perturbative method breaks down as $\sigma$ approaches unity. Note that $G(\nu) = 0$ for $\nu \lesssim -1$ in Figures 1 to 3 simply reflects the fact that corresponding critical density (3.5) becomes negative, and therefore the regions are not physically interesting.

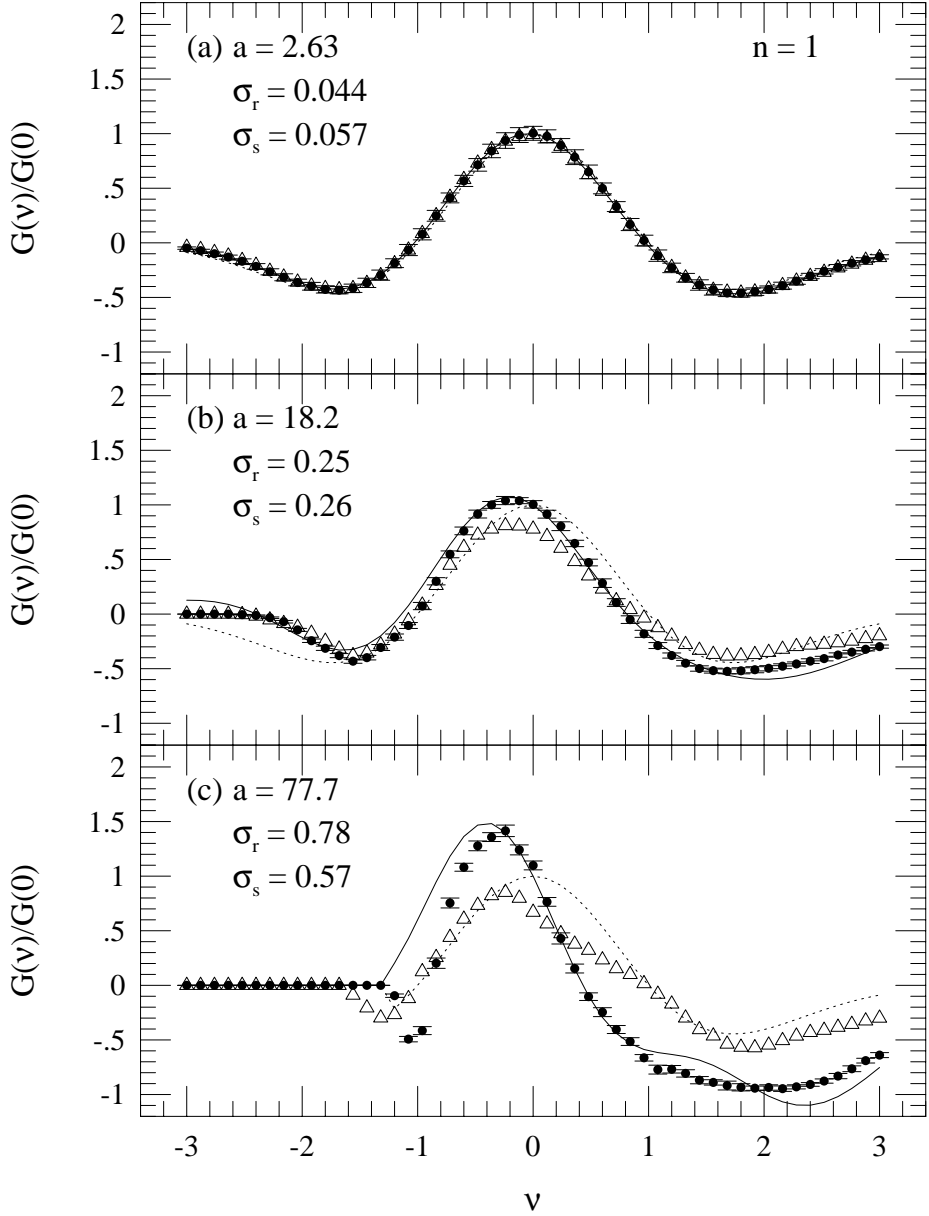

Figure 1: Normalized genus curves for $n = 1$ power-law model ($\Omega_0 = 1$, $\lambda_0 = 0$) in real space, $G^{(r)}(\nu)/G^{(r)}(0)$ (filled circles), and in redshift space, $G^{(s)}(\nu)/G^{(r)}(0)$ (open triangles). The dotted curve corresponds to the theoretical prediction for random-Gaussian field (2.3), while the solid curve takes account of weakly nonlinear evolution (eq.[2.7]). The Gaussian window function with the filtering length $R_f = L/25$ is used. The error bars represent the $1\sigma$ statistical error estimated with the 50 bootstrap resampling analyses. (a) $a = 2.63$, $\sigma_r = 0.044$, $\sigma_s = 0.057$; (b) $a = 18.2$, $\sigma_r = 0.25$, $\sigma_s = 0.26$; (c) $a = 77.7$, $\sigma_r = 0.78$, $\sigma_s = 0.57$.

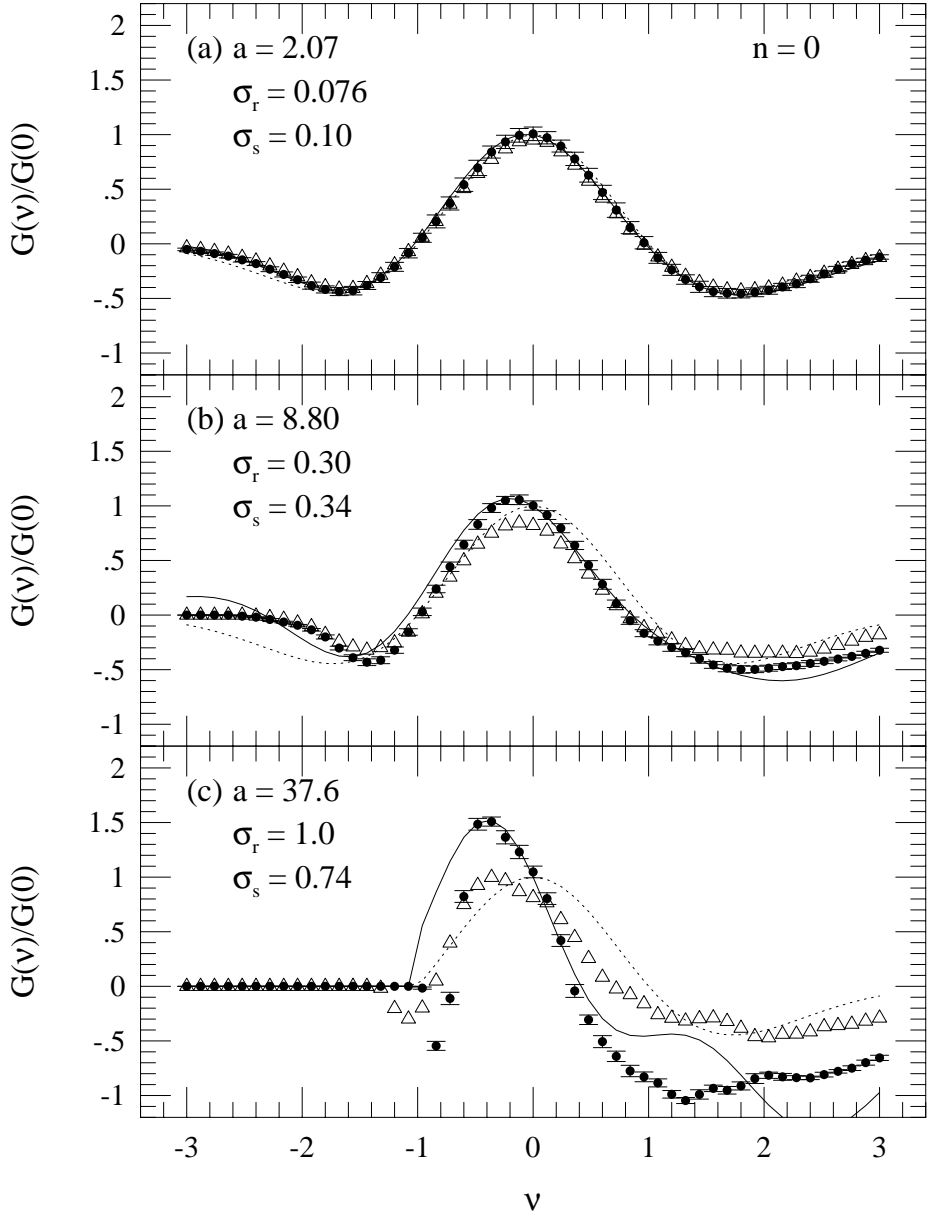

Figure 2: Same as Figure 1 for $n = 0$ power-law model. (a) $a = 2.07$, $\sigma_r = 0.076$, $\sigma_s = 0.10$; (b) $a = 8.80$, $\sigma_r = 0.30$, $\sigma_s = 0.34$; (c) $a = 37.6$, $\sigma_r = 1.0$, $\sigma_s = 0.74$.

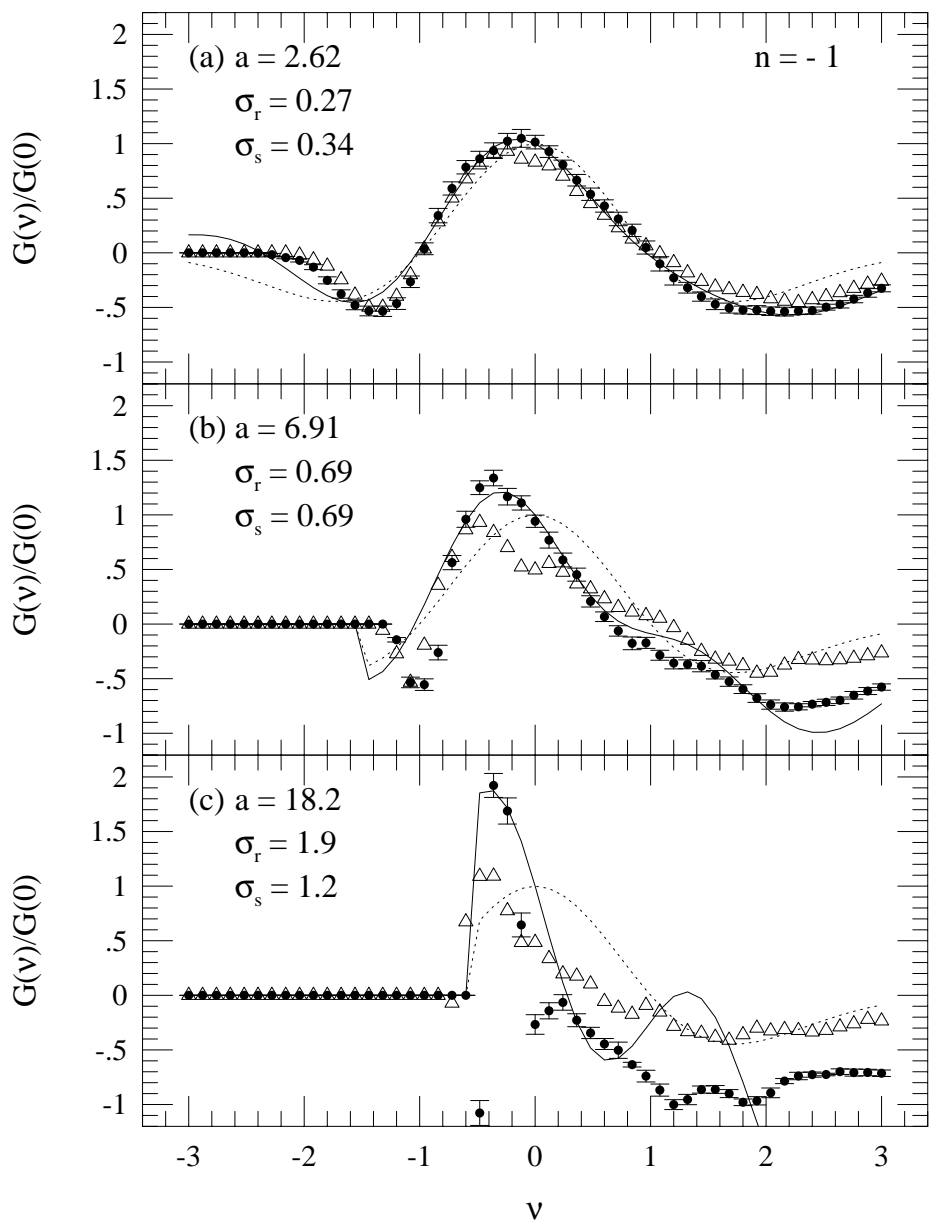

Figure 3: Same as Figure 1 for $n = -1$ power-law model. (a) $a = 2.62$, $\sigma_r = 0.27$, $\sigma_s = 0.34$; (b) $a = 6.91$, $\sigma_r = 0.69$, $\sigma_s = 0.69$; (c) $a = 18.2$, $\sigma_r = 1.9$, $\sigma_s = 1.2$.

Incidentally this agreement can be interpreted as yet another credibility of the $N$-body simulation method presented here. In fact, the simulations have been checked against theoretical predictions mainly via comparison of the evolution of the amplitude of the fluctuations $|\delta(k)|$ (i.e., $P(k)$ and $\xi(r)$; Suginohara et al. 1991; Suto 1993). Our present analysis is the first quantitative confirmation that $N$-body simulations faithfully reproduce the evolution of phases through the detailed comparison with the perturbation theory.

### 3.3. Redshift-space distortion

Matsubara (1996) showed that the genus curve $G(\nu)$ is fairly insensitive to the redshift-space distortion (§2.3). This tendency was also noticed earlier in numerical simulations by Melott, Weinberg & Gott (1988). Let us examine this point in details with our simulation results. First note that the genus curve in redshift space (open triangles in Figs. 1 to 3) is normalized by $G^{(r)}(0)$ in *real space* for the proper comparison with results in real space; thus their ratio is independent of the normalization factor $G^{(r)}(0)$.

As predicted in linear theory (§2.3), the redshift contamination is quite small where the rms fluctuation in real space $\sigma_r(R_f)$ is small. Linear theory predicts (Kaiser 1987) that the fluctuations in real and redshift spaces, $\sigma_r$ and $\sigma_s$, should satisfy

$$\frac{\sigma_s}{\sigma_r} = \sqrt{1 + \frac{2}{3}f + \frac{1}{5}f^2}. \tag{3.8}$$

Results shown in the upper panels of Figures 1 to 3 correspond to $\sigma_s/\sigma_r = 1.30$ ($n = 1$), $1.32$ ($n = 0$), and $1.26$ ($n = -1$) while equation (3.8) predicts $\sqrt{28/15} \sim 1.37$. Similarly $G^{(s)}(\nu)/G^{(r)}(\nu) = 39\sqrt{69}/343 \sim 0.94$ is predicted in linear theory (eq.[2.16]), while our results suggest $0.99$ ($n = 1$), $0.94$ ($n = 0$), and $0.82$ ($n = -1$).

The redshift-space distortion becomes noticeable in weakly nonlinear regimes, and tends to suppress the overall amplitude, i.e., $|G^{(s)}(\nu)/G^{(r)}(\nu)|$ more strongly than the prediction in linear theory (eqs.[2.16] to [2.19]). In a fully nonlinear regime $\sigma_r \sim 1$, $G^{(s)}(\nu)$ seems to approach the random-Gaussian prediction (dotted curves in Figs. 1 to 3). This feature originates from the fact that the velocity dispersions in virialized clusters are large and act effectively as the extra smoothing of the density field if computed in redshift space. This observable effect should be kept in mind in examining the primordial random-Gaussianity from the galaxy redshift surveys in limited volume size; obviously large homogeneous samples are important even in this respect.

Before closing this subsection, let us comment on the redshift distortion due to the cosmological expansion or general relativistic effect which becomes important as the sample depth of the galaxy redshift survey increases. In the Friedmann–Lemaître model, the comoving distance to the object at

$z$ is given by

$$d_C = \frac{c}{H_0} \begin{cases} (\Omega_0 + \lambda_0 - 1)^{-1/2}\sin(\chi\sqrt{\Omega_0 + \lambda_0 - 1}), & \text{for } \Omega_0 + \lambda_0 > 1 \\ \chi, & \text{for } \Omega_0 + \lambda_0 = 1 \\ (1 - \Omega_0 - \lambda_0)^{-1/2}\sinh(\chi\sqrt{1 - \Omega_0 - \lambda_0}), & \text{for } \Omega_0 + \lambda_0 < 1 \end{cases} \qquad (3.9)$$

where

$$\chi \equiv \int_0^z \left[\Omega_0(1+z)^3 + (1 - \Omega_0 - \lambda_0)(1+z)^2 + \lambda_0\right]^{-1/2} dz. \qquad (3.10)$$

For $z \ll 1$, $d_C$ is approximately given by

$$d_C = \frac{cz}{H_0}\left[1 + \frac{2\lambda_0 - \Omega_0 - 2}{4}z + O(z^2)\right]. \qquad (3.11)$$

Therefore the deviation from the simple linear Hubble law becomes appreciable even at relatively low $z$; at $z = 0.1$ ($\sim 300h^{-1}$Mpc), the *cosmological redshift-space distortion* becomes $-7.5\%$ in the Einstein-de Sitter model. Thus even for redshift surveys extending up to $z = 0.1$, this systematic effect dominates the statistical peculiar velocity effect ($\sim 3\%$ for $v = 1000$km/sec at $z = 0.1$, for instance). Although one can compute the genus of the observed sample in $d_C$ space rather than in $z$ space using equation (3.9) directly, it is not clear how the analytic expression (2.7) which neglect the cosmological evolution (or general relativity) can be compared with such results. In any case the result should be sensitive to the assumed set of $\Omega_0$ and $\lambda_0$. Note, however, that this problem is not specific to the genus statistics, but should be taken into account in the two-point and higher-order correlation analyses as well.

### 3.4. Smoothing effect of nonlinear distribution

The normalized genus curve $G(\nu)/G(0)$ in a weakly nonlinear regime (eq.[2.7]) is characterized only by the amplitude of the density fluctuations $\sigma$. In reality, however, $\sigma$ is dependent on both the epoch (or the expansion factor $a$) and the filtering length $R_f$ in smoothing the data. Thus it is possible that two different sets of $a$ and $R_f$ yield the same value of $\sigma$. Since the nonlinear evolution and the smoothing operation do not commute, it is of interest to ask whether those two different realizations give the same genus curve. In other words, to what extent can one suppress the strongly nonlinear contamination simply by smoothing the data with large filtering length ?

To answer this, we tried to find different sets of $a$ and $R_f$ which gives the same value of $\sigma$ for $n = 0$ power-law model. The resulting $G(\nu)/G(0)$ curves are shown in Figure 4a. This panel corresponds to the rms fluctuation amplitude $\sigma_r = 0.08$, and shows that the genus curves at later epochs and with larger $R_f$ deviate from the theoretical Gaussian prediction. In order to see the significance of this result, we have to check to what extent the error bars estimated from the bootstrap method is reliable, and also how the large filtering length affects the computation of the genus for samples in the finite volume.

For those purposes we performed two experiments. The first experiment divided the original cube into eight subcubes of half the boxsize, $L/2$, and computed directly the mean and the variance of the genus for eight subsamples *artificially* assuming the periodic boundary condition for each subcube. Then we estimate the mean and the variance for the original cube by averaging the means and adding the variances of the eight subsamples, respectively. The result is plotted in open circles with error bars (1 standard deviation) in Figure 4b. This should be compared with the mean (filled circles) and the $1\sigma$ error on the basis of the 50 bootstrap resampling analyses presented throughout this paper. Although the mean values derived from subsamples seem somewhat noisy, they are consistent with the bootstrap estimate and the perturbation theory (solid curve) within the error bars. In particular the two methods lead to almost the same error bars. This implies that the bootstrap estimate of the genus curve is reliable in the present context.

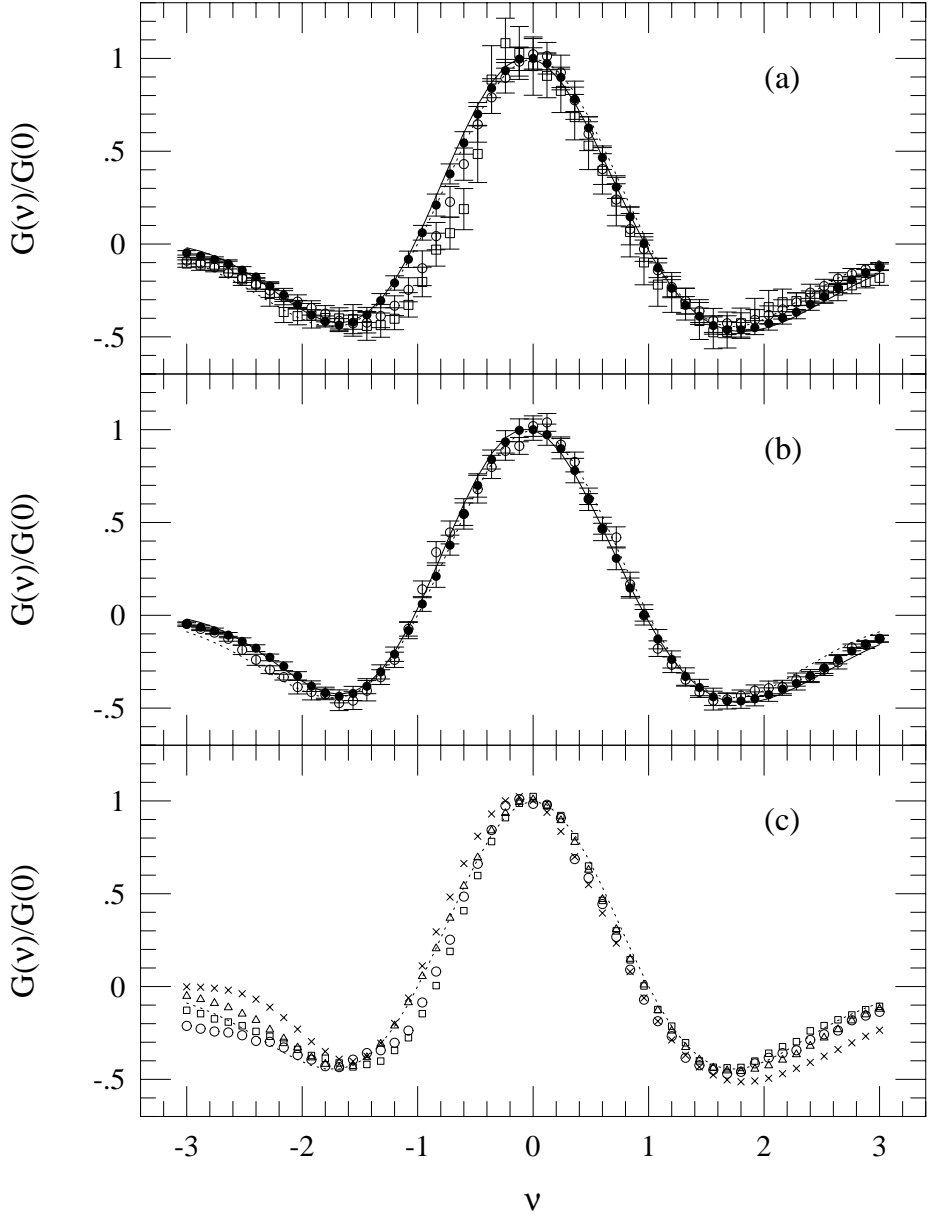

Figure 4: Smoothing effect on genus curves. (a) genus curves for $n = 0$ power-law models which have the same rms fluctuation amplitude $\sigma_r$ but for a different set of expansion factor $a$ and the Gaussian smoothing length $R_f$. The dotted and solid curves indicate the corresponding random-Gaussian (eq.[2.3]) and weakly nonlinear (eq.[2.7]) predictions for $\sigma_r = 0.08$. $(a, R_f/L, \sigma_r) = (2.07, 0.039, 0.0801), (4.27, 0.063, 0.00805)$, and $(8.80, 0.098, 0.0806)$ plotted in filled circles, open circles, and open squares, respectively. (b) Comparison with the means and error bars for $n = 0$ power-law model estimated from the bootstrap resampling method (filled circles) and from eight subsamples (open circles) for $(a, R_f/L, \sigma_r) = (2.07, 0.039, 0.0801)$. The dotted curve corresponds to the theoretical prediction for random-Gaussian field (2.3), while the solid curve takes account of weakly nonlinear evolution (eq.[2.7]). (c) Effect of the filtering length on the computation of genus curves for $n = 0$ power-law model at $a = 2.07$. $R_f/L = 0.02$) (crosses; $\sigma_r = 0.20$), 0.04(triangles; $\sigma_r = 0.076$), 0.08(squares; $\sigma_r = 0.027$), and 0.15 (circles; $\sigma_r = 0.0096$).

The second experiment checks the effect of increasing filtering length. Figure 4c shows results of $n = 0$ power-law model at $a = 2.07$ for $R_f/L = 0.02$ (crosses), 0.04(triangles), 0.08(squares), and 0.15 (circles). Since $\sigma_r$ decreases with increasing $R_f$, the resulting genus curves should approach the random-Gaussian prediction (dotted curve) for larger $R_f$. In fact this is basically the case for $\nu \gtrsim 0$, but not for $\nu \lesssim 0$. This unphysical result would be ascribed to the effect of the finite volume size of the present simulation; as $R_f$ approaches the boxsize $L$, the number of independent sampling volumes is reduced and the genus curve becomes more sensitive to the power of the large wavelength mode in one particular realization of our simulation models. Therefore the tendency shown in Figure 4a seems to be largely explained by the finite volume effect. In order to answer more definitively the question we addressed in this subsection, we need much larger simulations and/or many independent realizations which should be deferred to a future work.

Before closing the section, let us emphasize that the weakly nonlinear theory and numerical simulations are useful in a complementary manner to probe the statistics of the primordial density field using the genus curve; the former provides reliable, albeit with limited applicability, and definite predictions while the latter can in principle take into account fairly realistic effects including the higher-order nonlinear contribution, redshift-space distortion, sample-to-sample variation, and the finite volume effect of the observational sample. For specific predictions of the $G(\nu)/G(0)$ based on relatively realistic cosmological scenarios, we plot the results for LCDM model in Figure 5. The filtering length $R_f$ is $6h^{-1}$Mpc. If galaxies trace mass, $a = 6$ corresponds to the present epoch ($z = 0$). Thus $a = 4$ and 5 correspond to $z = 0.5$ and 0.2, respectively. The reasonable agreement between simulation and theory confirms the validity of the weakly nonlinear formula in $\Omega_0 < 1$ and $\lambda \neq 0$ models. Note that the results shown in Figures 1 to 4 are in the Einstein - de Sitter universe ($\Omega_0 = 1$ and $\lambda = 0$). The disagreement at a somewhat quantitative level, especially at $a = 6$, can be explained by the combined effects of the relatively large $\sigma_r$ which limits the validity of the weakly nonlinear formula (2.7), statistical fluctuations, and the significant departure from the random-Gaussian field *before smoothing* as mentioned above.

## 4.   DISCUSSION AND CONCLUSIONS

Genus statistics has been known as an important measure to quantify the topology of the large-scale structure in the universe. Its practical ability in constraining theoretical models, however, was limited mainly for two reasons; one is due to the fact that the currently available data are insufficient to unambiguously determine the genus curve from the observations, and the other is the lack of the theoretical prediction which properly takes account of the nonlinear gravitational effect. Actually the above two are closely related. In order to obtain statistically robust estimate of the genus from the limited spatial extent of the current observations, one has to use a relatively small filtering length in smoothing the data. Thus nonlinear gravitational effect becomes important. This difficulty was largely overcome with the analytic formula (Matsubara 1994b) to describe the genus in a weakly nonlinear regime. In addition the wide galaxy redshift surveys available in near future like Sloan Digital Sky Survey will improve the former.

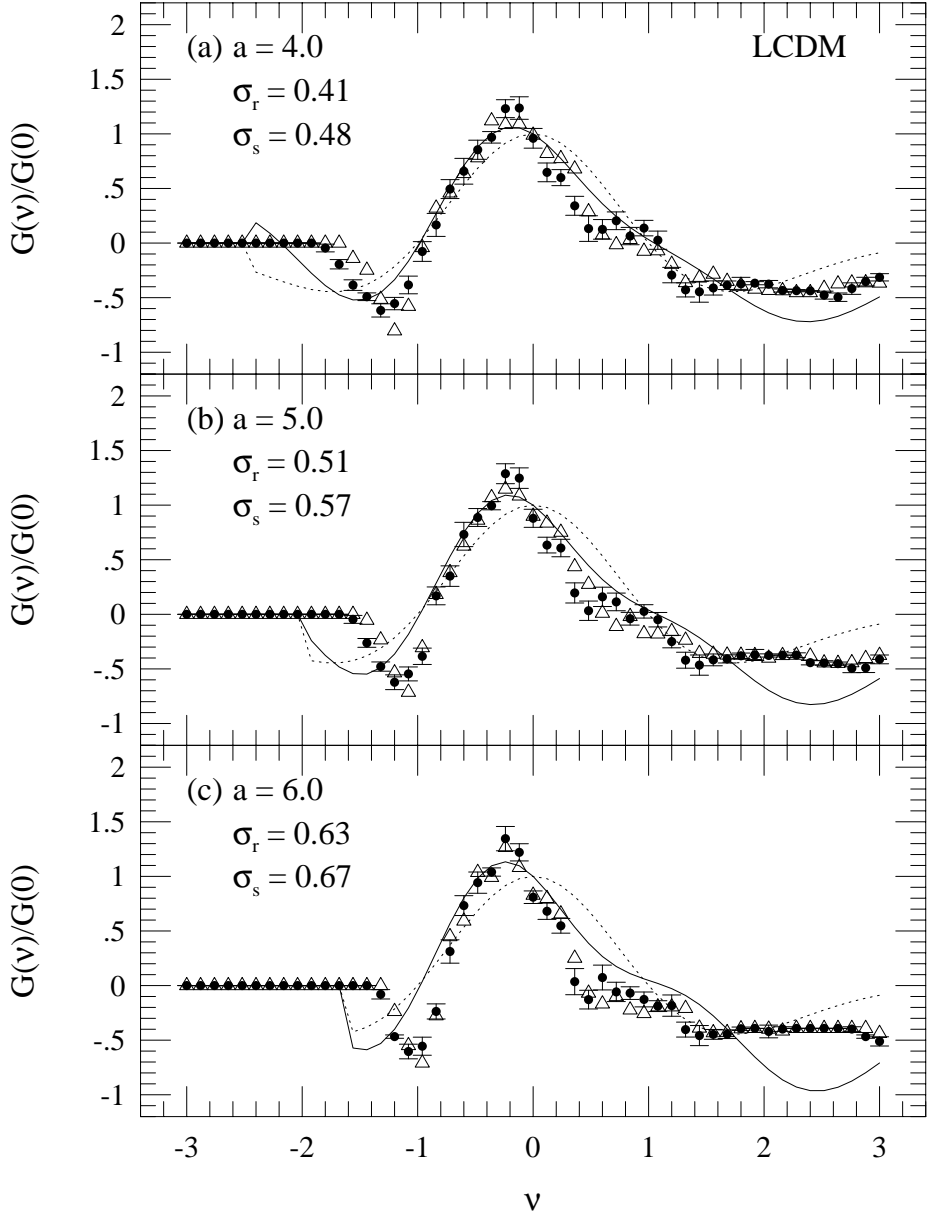

Figure 5: Same as Figure 1 for low-density cold dark matter model ($\Omega_0 = 0.2$, $\lambda_0 = 0.8$, $h = 1.0$). The adopted Gaussian filtering length $R_f$ corresponds to $6h^{-1}$Mpc (comoving). The top-hat smoothed rms mass fluctuation at $a = 6$ is unity at $8h^{-1}$Mpc. (a) $a = 4$, $\sigma_r = 0.41$, $\sigma_s = 0.48$; (b) $a = 5$, $\sigma_r = 0.51$, $\sigma_s = 0.57$; (c) $a = 6$, $\sigma_r = 0.63$, $\sigma_s = 0.67$.

This paper examined the validity of the formula using the large $N$-body simulations. In doing so we computed the genus curve in more realistic contexts including the nonlinear effect beyond the perturbative method, smoothing effect of the particle data in fully nonlinear regime, and the redshift-space distortion. We find that our simulation results are basically in good agreement with the weakly nonlinear formula derived by Matsubara (1994b), and that the redshift-space distortion on genus statistics is small in linear and weakly nonlinear regimes. Therefore we conclude that genus statistics, especially the normalized genus curve $G(\nu)/G(0)$, is an important tool to test the random-Gaussianity of the primordial density field if weakly nonlinear theory and direct numerical simulations are appropriately combined for the proper comparison with the real observations. Nevertheless there remain other potentially important questions concerning the genus statistics which include the selection effect in the magnitude-limited samples, biasing effect, the redshift-space distortion not only due to the peculiar velocity field of galaxies but also due to the cosmological expansion, and the finite volume effect. Unfortunately a realistic evaluation of these effects requires much larger simulation data and we should defer such an analysis to future work.

In summary, the genus curve $G(\nu)$ is a useful measure of galaxy clustering which is complementary to the two-point correlation function $\xi(r)$ in many respects; the normalized genus curve $G(\nu)/G(0)$ mainly quantifies the topology or the phase of the large-scale structure and is fairly insensitive to the underlying fluctuation spectrum. The latter is directly related to $\xi(r)$ which, on the contrary, contains no information of the phases $\phi(k)$. Once $G(\nu)/G(0)$ in a weakly nonlinear regime is determined, one can test the random-Gaussianity of the primordial density field, but it is almost independent of the underlying model and cosmological parameters (Matsubara 1994b ; see also §2.2). On the other hand, $\xi(r)$ is best suited to describe the galaxy clustering in fully nonlinear regime, and can constrain nature of dark matter, $\Omega_0$, and $\lambda_0$ among others by comparison with numerical simulations (Davis et al. 1985; Suto 1993). With future data of the wide survey of galaxy redshift, $G(\nu)/G(0)$ will enable us to directly check the random-Gaussian paradigm of the origin of the large-scale structure in the universe.

We thank David Weinberg for providing us the routines to compute genus curve from numerical data, and for stimulating discussions. We also thank an anonymous referee for pointing out a finite volume effect in computing the genus curve which significantly improves discussion presented in §3.4. T.M. gratefully acknowledges the fellowship from the Japan Society of Promotion of Science. This research was supported in part by the Grants-in-Aid by the Ministry of Education, Science and Culture of Japan (0042, 05640312, 06233209, 07740183, 07CE2002).